\documentclass[twocolumn,showpacs,preprintnumbers,amsmath,amssymb]{revtex4}
\usepackage{graphicx}% Include figure files
\usepackage{dcolumn}% Align table columns on decimal point
\usepackage{bm}% bold math

\begin{document}

\preprint{APS/123-QED}

\title{Hanbury Brown and Twiss correlations in atoms scattered from colliding condensates}

\author{Klaus M\o lmer}
\affiliation{Lundbeck Foundation Theoretical Center for Quantum
System Research, Department of Physics and Astronomy, University of
Aarhus, DK-8000 \AA rhus C, Denmark}
\author{A. Perrin, V. Krachmalnicoff, V. Leung, D. Boiron, A. Aspect, C.I. Westbrook}
\affiliation{Laboratoire Charles Fabry de l'Institut d'Optique, CNRS, Univ Paris-sud\\
Campus Polytechnique, RD128, 91127 Palaiseau, France}

\date{\today}

\begin{abstract}
Low energy elastic scattering between clouds of Bose condensed atoms
leads to the well known s-wave halo with atoms emerging in all
directions from the collision zone. In this paper we discuss the
emergence of Hanbury Brown and Twiss coincidences between atoms
scattered in nearly parallel directions. We develop a simple model
that explains the observations in terms of an interference involving
two pairs of atoms each associated with the elementary s wave
scattering process.

\end{abstract}


\maketitle

\section{Introduction}

In a number of experiments, Bose-Einstein condensates have been prepared to collide with each other
with well defined collision energies and momenta. At the microscopic level, when two particles with
equal mass and opposite velocities collide in an s-wave collision, the collision partners will
propagate away from each other with the same probability amplitude in all directions, but their
individual momenta are correlated in opposite directions, as their total center-of-mass momentum is
conserved in the collision process. In experiments with colliding condensates, the scattering into all
directions has been clearly observed as a so-called s-wave halo of scattered particles\cite{collisions}. The observation of  pair correlations of particle leaving the
collision region back-to-back, see Fig. 1, requires efficient detection of all momentum components of
individual atoms, and this correlation has recently been observed as a significant coincidence signal
in a collision experiment with Bose-Einstein condensates of metastable atomic helium \cite{orsayexp}.
The same experiment also observed an increased coincidence of particles scattered into \emph{the
same} direction. This phenomenon is due to the bosonic nature of the particles and to the fact that
several independent scattering processes occur simultaneously.
\begin{figure}
  \includegraphics[width=6cm]{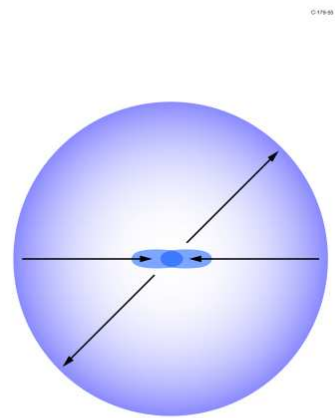}\\
  \caption{Diagrammatic representation of two condensates colliding and giving rise to an s-wave
  halo of scattered particles. Particles are scattered pairwise back-to-back.}\label{fig1}
\end{figure}

We shall present a theoretical analysis of this Hanbury Brown and Twiss correlation phenomenon, aiming
at a simple model which explains its qualitative and quantitative character in experiments. It is
important to emphasize that the appearance of atoms moving in the same direction is not compatible
with momentum conservation in a single collision process of two counterpropagating atoms, and our
discussion will, indeed, refer to effects that rely on a many-body treatment of the collision of
larger ensembles. In order to get physical insight, we will separate the problem in two: first, we
treat independent pairwise collisions in which the collisional interaction gives rise to pairs of
scattered atoms, and second, we treat the evolution of the many-body state describing the ensemble of
scattered atoms neglecting the interactions in order to apply analytic methods. The validity of
this separation and means to improve the theory, if necessary, will be discussed.

In Sec. II, we present a full second quantized description of the collision of a large number of
identical bosons. We shall write down the second quantized many-body Hamiltonian, and discuss how the elementary processes of interest relate to the different terms in this
Hamiltonian. We treat the case where all the atoms initially populate two counter propagating single
particle states which are only weakly depleted by the collisions, and which will hence serve as
c-number field sources for creation of pairs
%due to collision of atoms out of the macroscopically
%populated modes.
This is in analogy with the quantum optics treatment of spontaneous
four-wave mixing, where photon pairs are generated from the interaction of two incident laser beams,
described by classical electromagnetic waves. In Sec. III, we discuss the Bogoliubov transformation
which provides a very accurate approximation to the time evolution of the system. We shall not,
however, apply this transformation in a quantitative treatment, but rather show that its
formal structure already predicts the collinear (Hanbury Brown and Twiss) correlation and motivates a quite general
analytical Ansatz for the quantum state of the scattered atoms. In Sec. IV, we shall consider the
leading two- and four-atom terms in an expansion of the quantum state of scattered atoms, and show
that they hold the key to the observed Hanbury Brown and Twiss correlations. In Sec. V, we shall use
energy conservation and phase matching considerations to motivate a simple analytical model, from
which we show that the coincidence of scattered particles in the same direction, though a
many-body effect, can be understood quantitatively from the properties of the simple two-atom
scattering wave function. Sec. VI concludes the paper with a discussion of the insights offered by our
analysis.

\section{Colliding Bose-Einstein condensates}

The Hamiltonian,
\begin{equation} \label{fullH}
H=\int d\vec{r} \hat{\Psi}^{\dagger}( \vec{r})\left( -\frac{\hbar^2}{2m}\Delta +
V(\vec{r})+\frac{g}{2}\hat{\Psi}^{\dagger}(\vec{r})\hat{\Psi}(\vec{r})\right) \hat{\Psi}(\vec{r}),
\end{equation}
with field operators obeying the bosonic commutator relations
$[\Psi(\vec{r}),\Psi^{\dagger}(\vec{r}')]=\delta(\vec{r}-\vec{r}')$, gives a good description of
bosons interacting at low collision energies via a short range potential. The interaction is
represented by a delta-function interaction term with strength $g$, proportional to the s-wave
scattering length. The atoms may be subject to a wide range of trapping or guiding potentials
$V(\vec{r})$, or they may propagate freely $(V=0)$, and the initial state of the system may be
specified according to experimental preparation procedures to describe for example a single condensate
or several macroscopically populated components. We are interested in the situation, where two
condensates with well defined opposite momenta, and hence with relatively large spatial extent
propagate towards each other. The conventional second quantized Hamiltonian fully describes the
problem, and a Monte-Carlo type simulation of the dynamics \cite{Drummond, Deuar, KheruntsyanPriv}, and
perhaps even simpler simulation approaches based on truncated Wigner function expansions
\cite{Sinatra,GardinerPriv}, may solve this problem in full generality  by full 3D propagation of
stochastic Schr\"odinger type equations.

We assume that elastic collisions occur with a sufficiently small cross section that the colliding
condensates are only weakly depleted due to the collision term in (\ref{fullH}).  The Hamiltonian has
terms describing the kinetic energy and the potential energy of atoms moving in the external potential
and finally, a term describing the mean field repulsive or attractive potential due to the other atoms
of the colliding condensates. But the product of two creation and two annihilation operators in the
interaction term does not only read as density dependent correction to the potential energy in the
Gross-Pitaevskii equation: the product of two creation operators may also cause the creation of a pair
of atoms with momenta entirely different from the incident ones, extracted consistently from the
condensates by the product of annihilation operators. The  pairs of atoms ``created" in the scattering
process are the ones that are detected as the s-wave halo around the condensate collision region in
Fig. 1.

We can think of each point in the collision zone as a point source
for a pair of initially close atoms (atoms only collide at short
range), which are subsequently separated by free
propagation, perturbed by the interaction with the condensate
components. This propagation, together with the coherent addition of
pair amplitudes originating from the entire collision zone leads to
a complicated many-body entangled state, but energy conservation,
imposed after sufficiently long interaction time, and momentum
conservation, imposed by phase matching, serves to justify our
simpler model, described below.

\section{Bogoliubov approximation}

If the original condensates are only weakly depleted by the scattering, we may expand the Hamiltonian
to second order in operator terms and get linear Heisenberg equations of motion that may be
solved and tell precisely which Bogoliubov transformation exactly solves the problem.  This approach
was followed in Refs. \cite{Rz1,Rz2,Rz3}. The Bogoliubov solution yields an expression for the atomic
annihilation operators at any given time, expressed as a linear combination of the annihilation and
creation operators at time zero, where the initial state is assumed to be known (incident condensate wave
functions, no scattered atoms). The mean atom number and any higher order correlation function of the
field can therefore be expressed algebraically in terms of the expansion coefficients of the
Bogoliubov transformation and the known vacuum expectation values of field operator products.

Even though the full many problem has thus been reduced to partial
differential equations of a complexity comparable to the single
particle Schr\"odinger equation, one still has to solve time
dependent wave equations in three spatial dimensions. Here we shall
demonstrate some properties of the solution that
follow by purely analytical arguments, i.e., without access to the
precise solution.

Although obviously related, the use of the Bogoliubov
transformation here is different from the Bogoliubov approximation
used to identify low-lying, collective excitation modes in a
condensate. The analysis rather follows the philosophy of squeezed
light generation with optical parametric oscillators in quantum
optics, where the Bogoliubov method is used to diagonalize a
multi-mode Hamiltonian with pair creation and annihilation operators
\cite{squeezedlight}, see also \cite{Uffemol,Uffe07}.

Since the initial state of the atomic scattering modes (momentum components) of interest is the vacuum
state, which has a Gaussian (Wigner) probability distribution  for the multi-mode field variables, and
the Bogoliubov transformation is linear in field operators, the state will, independently of the
precise form of the transformation, at all later times be a Gaussian with vanishing mean field
expectation value \cite{ekertknight}. If we restrict the analysis to a single final momentum state
(mode), by a partial trace over all other modes, the state of this mode is also a Gaussian state with
vanishing mean amplitude. It is thus fully characterized by the second moments of the hermitian linear
combinations $q_{\vec{k}}\equiv \left(\hat{\Psi}^\dagger(\vec{k})+
\hat{\Psi}(\vec{k})\right)/\sqrt{2},\ p_{\vec{k}}=i\left(\hat{\Psi}^\dagger(\vec{k})-
\hat{\Psi}(\vec{k})\right)/\sqrt{2}$ of the field operators. We now wish to establish that our
Gaussian distribution is symmetric, \emph{i.e.}, Var$(q_{\vec{k}}$)=Var$(p_{\vec{k}}$). This indeed
follows if the "anomalous" moments $\langle \hat{\Psi}^\dagger(\vec{k})^2\rangle=\langle
\hat{\Psi}(\vec{k})^2\rangle=0$, \emph{i.e.}, if there is no coherence between states differing by two
atoms propagating in the given direction. We now apply the physical argument, that the collisional
Hamiltonian does not produce such coherence, since the collision process can only produce pairs of
atoms propagating in opposite directions, and states, e.g, with zero and two atoms with momentum
$\vec{k}$ must also contain zero and two atoms with momentum around $-\vec{k}$. The anomalous moments
vanish due to the orthogonality of these parts of the wave function.

It is well known in quantum optics, that a symmetric Gaussian state is equivalent to an incoherent
mixture of number states with exponential number distribution, also known as a thermal state with the
density matrix \cite{squeezedlight},
\begin{equation} \label{onemode} \rho_{1}=(1-|t|^2)\sum |t^2|^n
|n\rangle\langle n|.
\end{equation}
The state conditioned upon detection, and annihilation, of a single
particle reads,
\begin{equation} \label{cond}
\rho_c = \nu \hat{a}\rho_1\hat{a}^+ = \frac{(1-|t|^2)^2}{|t|^2}
\sum_n |t^2|^n n |n-1\rangle\langle n-1|,
\end{equation}
where $\nu$ is a normalization constant. A straightforward calculation shows that this state has
precisely twice as many bosons on average as (\ref{onemode}), and hence that the probability to detect
two bosons by a low efficiency detector is twice the square of the single quantum detection
probability. It thus follows that the coincidence counting rate for observing two atoms leaving the
collision zone in the same, narrowly defined, direction, $\langle
\hat{\Psi}^{\dagger}(\vec{k})\hat{\Psi}^{\dagger}(\vec{k})\hat{\Psi}(\vec{k})\hat{\Psi}(\vec{k})\rangle
$ is twice the square of the mean counting rate, and twice the coincidence rate for seeing atoms in
two unrelated directions.

Without performing any calculations, we therefore understand qualitatively the observed coincidences
observed in the experiments \cite{orsayexp} as the direct consequence of the thermal counting
statistics (Gaussian quadrature distribution) of the output flux in all scattering directions. This is
the famous Hanbury Brown and Twiss effect \cite{hbt1,hbt2,hbt3,mandel} observed originally as photon
bunching in chaotic light resulting from the addition of the contributions of many incoherent
emitters. In order to provide a natural estimate of the HBT momentum correlation function, one could
develop the field correlations by solution of the linear Bogoliubov-de-Gennes equations for the
problem \cite{Rz1,Rz2,Rz3} which by the corresponding linear transformation of operators provides the
first and second order momentum correlation functions and hence the momentum range within which the
bunching effect takes place. Here, we will rather keep track of the binary scattering states, and in
particular of the counterpropagating partners, which will give us an alternative and very useful
physical interpretation of the effect.
%
%A full multi-mode Hamiltonian which is quadratic in annihilation and
%creation operators can by Bogoliubov diagonalization be brought on
%an exact block-diagonal form, involving modes and pairs of modes,
%which are either single-mode squeezed or pair correlated with exact
%number matching of the two modes \cite{Uffe07}. Since atoms are
%created in pairs with opposite momenta, according to the above
%argument, we do not expect any amplitude squeezing in single modes,
%but due to the finite momentum width of the colliding condensates,
%the scattered particles will not have definite total momentum, and
%the Hamiltonian is not diagonal in the momentum state basis.

The Bogoliubov transformation of field operators is equivalent to a multi-mode unitary squeezing
operation \cite{squeezedlight}, which is indeed nothing but the time evolution operator of a
Hamiltonian with quadratic terms in field creation and annihilation operators. Such an operator can be
ordered as a product of three exponentials \cite{normorder,iwop}: one involving a sum of products of
pairs of creation operators, one involving a sum of products of creation and annihilation operators
and one involving a sum of products of pairs of annihilation operators. When acting on the initial
vacuum state vector, only the unit term of the series expansion of the latter two exponentials
contribute, and the state can therefore be written in terms of a quadratic form of creation operators
of atoms, e.g., in the momentum space representation,
\begin{equation}
\label{bog}
|\Psi> = N_{\Psi} \exp(\int d\vec{k}_1 d\vec{k}_2
\psi(\vec{k}_1,\vec{k}_2)\hat{\Psi}^{\dagger}(\vec{k}_1)\hat{\Psi}^{\dagger}(\vec{k}_2))|vac>.
\end{equation}
The function $\psi(\vec{k}_1,\vec{k}_2)$ generally depends in a
complex %difficult  (klaus you seem to have changed this back but it still sounds funny to me.)
manner on the dynamical
evolution. It is of course related to the scattering wave function of a single pair of atoms, and we
shall come back to this relationship in connection with the model studied in Sec. V of the paper. The
use of second quantization automatically yields the bosonic symmetry of our state, but in addition we
can require that the pair amplitude function obey the explicit exchange symmetry
$\psi(\vec{k}_1,\vec{k}_2)=\psi(\vec{k}_2,\vec{k}_1)$. For now, let us assume, that the propensity for
atoms to be scattered into opposite direction also implies that $\psi(\vec{k}_1,\vec{k}_2)$ takes
non-vanishing values for all directions of the scattered particles, but only if $\vec{k}_1 \sim
-\vec{k}_2$. The function $\psi(\vec{k}_1,\vec{k}_2)$ is not a normalized wave function: the larger
its amplitude the more particle pairs are created, and higher order terms of the exponential play more
and more important roles. The many-body state $|\Psi\rangle$ is normalized by the prefactor $N_{\Psi}$
in (\ref{bog}).

We now proceed to determine the density-density correlations of
atoms detected in two different directions, labeled by momentum
states $(\vec{k},\vec{k}')$, i.e., the expectation value
\begin{equation} \label{corr}
F(\vec{k},\vec{k}') \propto \langle
\hat{\Psi}^+(\vec{k})\hat{\Psi}^+(\vec{k}')\hat{\Psi}(\vec{k}')\hat{\Psi}(\vec{k})\rangle
\end{equation}

The state (\ref{bog}) is a Gaussian state, and by Wick's theorem \cite{louisell} this expectation
value can be written down in terms of only pair-expectation values. We shall address the contribution
from the four-atom component in the expansion of the exponential in (\ref{bog}), as this provides a
straightforward interpretation of the origin and the behavior of the atomic Hanbury Brown and Twiss
correlations.

\section{Two-atom and four-atom states}

The state (\ref{bog}) can be written explicitly,
\begin{eqnarray} \label{tall}
|\Psi\rangle & = & N_\Psi ( |vac>\nonumber \\
& + & \int d\vec{k}_1 d\vec{k}_2
\psi(\vec{k}_1,\vec{k}_2)\hat{\Psi}^+(\vec{k}_1)\hat{\Psi}^+(\vec{k}_2)|vac>\nonumber
\\
 & + & \frac{1}{2}(\int d\vec{k}_1 d\vec{k}_2
\psi(\vec{k}_1,\vec{k}_2)\hat{\Psi}^+(\vec{k}_1)\hat{\Psi}^+(\vec{k}_2))^2|vac>\nonumber
\\
 & + & ... ).
\end{eqnarray}
The zero order term is the vacuum state. The first order term is a two-atom state of atoms propagating
back-to-back, and the second order term of the series expansion of (\ref{bog}) is the four-atom state

\begin{equation} \label{psi4}
|\Psi_4\rangle\equiv (\int d\vec{k}_1 d\vec{k}_2
\psi(\vec{k}_1,\vec{k}_2)\hat{\Psi}^+(\vec{k}_1)\hat{\Psi}^+(\vec{k}_2))^2|vac>,
\end{equation}
which we will show accounts for the observed HBT effect. The squared pair creation operator in
(\ref{psi4}) can be expanded as a four-fold integral.  To obtain the correlation function
(\ref{corr}), we have to apply the product of the two annihilation operators on $|\Psi_4\rangle$ and
determine the squared norm of the resulting state,
\begin{equation} \label{fun}
F(\vec{k},\vec{k}') \propto
||\hat{\Psi}(\vec{k})\hat{\Psi}(\vec{k}')|\Psi_4\rangle||^2.
\end{equation}
Using the field commutator relations, we can shift the annihilation
operators to the right of all creation operators in (\ref{fun}).
This yields a total of 12 terms, which by relabeling and use of the
exchange symmetry can be reduced to a sum of three different
contributions,
\begin{eqnarray} \label{ops}
\hat{\Psi}(\vec{k})\hat{\Psi}(\vec{k}')|\Psi_4\rangle \propto \nonumber \\
\int d\vec{k}_1 d\vec{k}_2 \{\psi(\vec{k}_1,\vec{k}_2)\psi(\vec{k},\vec{k}') +
\psi(\vec{k}_1,\vec{k})\psi(\vec{k}_2,\vec{k}') \nonumber \\ +
\psi(\vec{k}_1,\vec{k}')\psi(\vec{k}_2,\vec{k})\}
\hat{\Psi}^+(\vec{k}_1)\hat{\Psi}^+(\vec{k}_2)|vac>.
\end{eqnarray}
and thus its squared norm:
\begin{eqnarray} \label{funfin}
F(\vec{k},\vec{k}') \propto && \int d\vec{k}_1 d\vec{k}_2
|\psi(\vec{k}_1,\vec{k}_2)\psi(\vec{k},\vec{k}')\nonumber
\\ + && \psi(\vec{k}_1,\vec{k})\psi(\vec{k}_2,\vec{k}') +
\psi(\vec{k}_1,\vec{k}')\psi(\vec{k}_2,\vec{k})|^2.
\end{eqnarray}

This is the main result of the paper. Dealing explicitly with the four atom component it is easy to
see what happens. In Fig. 2. we illustrate the case of detection of a particle pair in random
directions. Because, as noted below Eq.~(4), $\psi(\vec{k},\vec{k}')$ vanishes unless $\vec{k}$ and
$\vec{k}'$ are anti-parallel, the first term in Eq.~(10) only contributes if the detectors correspond
to opposite directions. For opposite or random directions such as in Fig. 2, there is also no cross term between the second two terms because one vanishes whenever the other is finite.
If $(\vec{k},\vec{k}')$ are nearly parallel, as illustrated in Fig.~3,
%If the detectors are not put in two opposite directions, the first term in
%(\ref{funfin}) vanishes and the expression (\ref{funfin}) becomes the integral of the sum of squares
%of the last two terms. There is no cross term because one component vanishes whenever the other is
%finite. For the case of nearly parallel $(\vec{k},\vec{k}')$, illustrated in Fig.3, the first term in
%(\ref{funfin}) vanishes, because $\psi(\vec{k},\vec{k}')$ is small for parallel arguments.
the last two terms evaluate the two different $\psi$-terms at the detector directions and at the
direction specified by the integration variables. This means that values of the integration variables
$\vec{k}_1,\vec{k}_2$ exist (opposite to the detector directions), where  both of the last terms in
(\ref{funfin}) contribute, namely if $\vec{k}_1$ and $\vec{k}_2$ are within the ``recoil cone" of both
detection directions $\vec{k}$ and $\vec{k}'$. For identical $\vec{k}$ and $\vec{k}'$ this
interference give precisely the factor 2 increase of coincidences compared to the case of random
directions. We also note that the enhanced coincidences occur within a solid angle specified precisely
by this "recoil cone". The next section develops our model one step further and carries out the
calculation for the special choice of a Gaussian Ansatz for the function $\psi(\vec{k}_1,\vec{k}_2)$.
\begin{figure}
  \includegraphics[width=6cm]{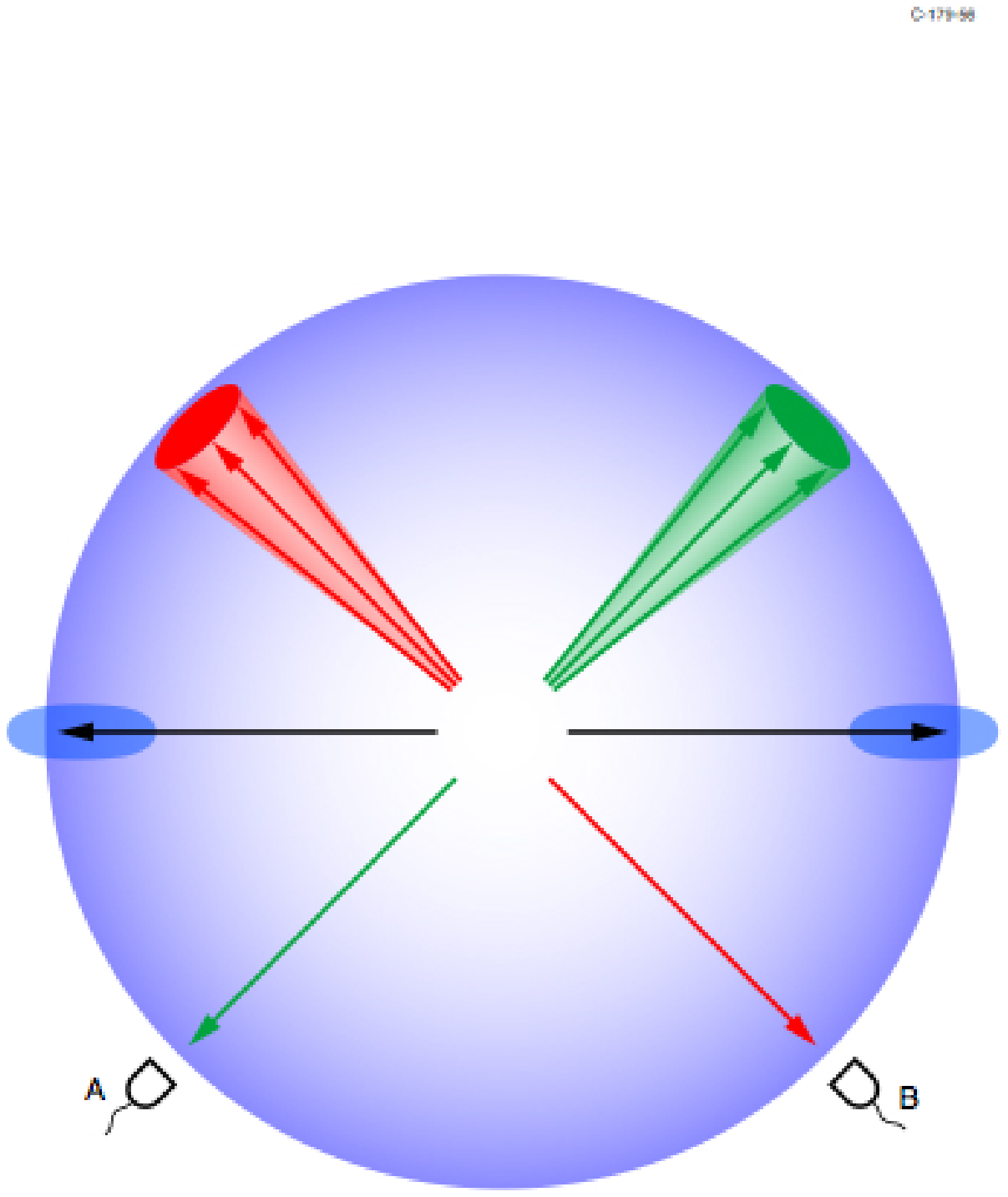}\\
  \caption{State of four atoms, scattered pairwise back-to-back. Atoms detected by detectors A and B
  in an arbitrary pair of directions have  partners recoiling in the opposite directions
  within a certain width imposed by the uncertainty on total and relative momentum of the atoms.
  The quantum state of the detected pair is obtained by a partial trace over the recoiling momentum
  components, and the coincidence counting yield in the detectors is just the product of the single detector
  count signals }\label{fig2}
\end{figure}

\begin{figure}
  \includegraphics[width=6cm]{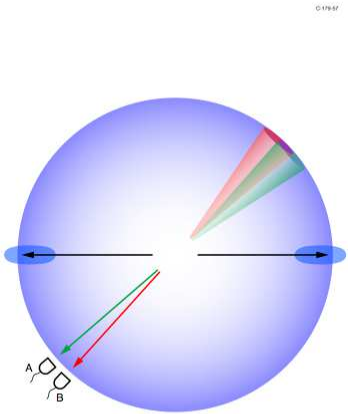}\\
  \caption{State of four atoms, with two atoms detected in nearly parallel directions. The detected atoms are not
  unambiguously identified with their recoiling partners
  if their momentum distributions are wide enough to overlap. This leads to an interference term in the coincidence counting
  yield in detectors A and B as expressed by the formal expression (\ref{funfin}).}\label{fig3}
\end{figure}

\section{Result for a simple Ansatz for $\psi(\vec{k}_1,\vec{k}_2)$}

As stated above, the Bogoliubov approximation to the actual state can be found numerically by solution
of linear wave equations. In this section, we shall rather take a simpler approach by making an Ansatz for the shape of the
function $\psi$ by appealing to the dynamics and the conservation laws valid in the bipartite collision
dynamics.

Energy conservation, which is effectively enforced during the temporal solution of the Schr\"odinger
equation, suggests that the atomic pair state $\psi(\vec{k}_1,\vec{k}_2)$ describes particles with the
same energy as the incident condensate particles. Momentum conservation suggests that they also have
the same total momentum as the colliding pair. Due to the finite size of the colliding clouds, this
does not strictly imply that the two atoms must have exactly opposite momenta. The finite size of the
collision zone implies a quantum mechanical momentum uncertainty, and if contributions from small
regions (with correspondingly large total momentum uncertainty) are added coherently the resulting
phase matching condition is not sharp if the entire collision zone has finite size.
There are therefore quantum fluctuations of both the
 modulus and direction of the momenta.
%There is therefore
%both some allowed quantum fluctuations of the magnitude of the momenta and of their components
%perpendicular to their mean directions. Although these fluctuations may be of similar size, the
%transverse ones are of most interest to us here as they are easily compared with the angular
%distribution of particles observed in the experiment. We restrict the analysis here to transverse
%components.
When the particles escape from the collision zone as illustrated in Fig. 1., they are also repelled by
the mean field interaction with the two condensate components. Here we will not try to describe
accurately the effect of this interaction, which is in any case small when atoms leave a condensate
after receiving an initial
%non vanishing momentum
kinetic energy large compared to the chemical
potential \cite{robins}. Note, however, that such
mean field repulsion has in fact turned out to be tractable in the atom laser output from a
condensate, where a generalized ABCD matrix formalism yields and analytical description of the
propagation \cite{orsayatomlaser}.

Define coordinates such that the nearly parallel $\vec{k},\vec{k}'$ of interest are close to the
negative $z$-direction. Their partners at $\vec{k}_1,\vec{k}_2$ must both be close to the positive
$z$-direction, and we shall assume the $z$-coordinates to be equal and opposite and only look at their
$x$ and $y$ components. Their widths are related to the wave functions of the colliding condensates,
both due to the amplitude of collisions out of these condensates and due to the mean field repulsion,
and they are thus in general anisotropic. We restrict for simplicity the integration to one transverse
coordinate (putting vector arrows on the arguments will yield the 2D result), and we assume that a
single pair is described by a wave function, where the wave function amplitude for the recoiling
partner has a bell shaped profile, that we for simplicity approximate by a Gaussian, centered at minus
the coordinates of the detected particle. The width of this wave function is parametrized by a
momentum width $K$ which thus represents both the momentum width of the colliding condensate particles
and the acceleration due to the mean field.

Assuming thus the last two terms in (\ref{funfin}) to be of such Gaussian shape, and ignoring the
first term which vanishes for the geometry studied, we can explicitly calculate the coincidence
signal:
\begin{eqnarray} \label{gauss}
F(\vec{k},\vec{k}') \propto &&\int dk_1 dk_2 \nonumber \\
&&|\exp(-((k_1+k)^2+(k_2+k')^2)/2K^2)\nonumber \\ +
&&\exp(-((k_1+k')^2+(k_2+k)^2)/2K^2) |^2\nonumber \\
\propto && 1+\exp(-(k-k')^2/2K^2),
\end{eqnarray}
where we recall that $k$ and $k'$ here refer to (small) transverse
coordinates of the detector directions with respect to a given axis,
i.e. $(k-k')$ is the radial momentum of the outgoing particles
multiplied with their mutual (small) angle in radians.

We recover the Hanbury Brown and Twiss correlations, and we observe
that the correlations persist for final state momenta within a
distance from each other of the order of the quantum mechanical
uncertainty of the total momentum of the atom pair escaping the
collision zone. This is in accord with our interpretation in terms
of the interference between the indistinguishable components
illustrated as the overlapping recoil cones in Fig.1 c), that leads
to the last term in (\ref{gauss}) depending on both $k$ and $k'$,
whereas the direct terms lose the $k$ and $k'$ dependence due to
the Gaussian integrals.

It is interesting to note, that Eq.(\ref{gauss}) follows from a two-state amplitude $\psi(k,k')
\propto \exp(-(k+k')^2/2K^2)$ for transverse momentum components of atoms propagating in nearly
opposite directions, and therefore the two atom component of Eq.(\ref{tall}) predicts a correlation of
atoms in opposite directions with the dependence $|\psi(k,k')|^2 \propto \exp(-(k+k')^2/K^2)$. The
Hanbury Brown and Twiss bunching thus occurs within a Gaussian width that is $\sqrt{2}$ times larger
than the range of correlation of recoiling atomic momenta. This prediction for the Gaussian wave
functions has been verified by more detailed analysis of the full 3D propagation,
\cite{KheruntsyanPriv}. One way to understand the broadening is to recognize that the density
dependence of the pair production mechanism results in a source which is spatially narrower than the
condensates themselves.

Although we have based our analysis on the four-atom component of the full many body states, we have
argued that a calculation based on the full state would yield the same results, and in particular that
the HBT correlation amounts to a factor of two in parallel directions while the correlation in
opposite directions is not limited by this factor. When multiple scattering is neglected, the complete
many-body problem is solved by the Bogoliubov-de-Gennes equations, and the resulting Gaussian/thermal
character of the many-body state is fully accounted for by the second moments. This does not imply,
however, that one would get the same quantitative results for scattering of few and many atoms. If the
normalized wave function for a single scattered pair in the case of a low scattering probability is
denoted $\chi(\vec{k}_1,\vec{k}_2)$, it may be reasonable to describe the collision process by the
effective Hamiltonian $H=\kappa\int d\vec{k}_1 d\vec{k}_2 \chi(\vec{k}_1,\vec{k}_2)
\hat{\Psi}^\dagger(\vec{k}_1)\hat{\Psi}^\dagger(\vec{k}_2) + h.c.$, where $\kappa$ is a coupling
strength. The unitary time evolution operator is the exponential of this operator multiplied by
$(t/i\hbar)$ or integrated over a suitable time interval. We note that this does not generally result
in an expression for $\psi(\vec{k}_1,\vec{k}_2)$ in (\ref{bog}) which is proportional to
$\chi(\vec{k}_1,\vec{k}_2)$. In the case of single mode squeezing, it is known that one must evaluate
the hyperbolic tangent function of the squeezing parameter to convert the squeezing operator to the
normal order form \cite{squeezedlight}, and in our general multimode case, normal ordering is
accomplished by evaluating the $tanh$ function of a matrix argument \cite{normorder,iwop}. For small
arguments, in the perturbative regime of spontaneous four wave mixing, $tanh$ is a linear function,
and we get the same momentum dependence. Outside the perturbative regime, we retain the factor 2
bunching effect by our general argument, but the precise shape of the correlation peak may be
modified.

\section{Discussion}

We have presented a simple interpretation of the observed Hanbury Brown and Twiss correlations
observed in the elastic scattering of Bose-Einstein condensates. We emphasize, that in order to make
quantitative predictions, it is necessary to make a more elaborate calculation of the pair formation
and the propagation of the atoms both in free space and in the regions where the mean field of the
condensate components act as a perturbing potential. Such a description is offered by the Bogoliubov
theory in Refs.\cite{Rz1,Rz2,Rz3}, and we note that \cite{Rz3} as well as \cite{Deuar} also provide numerical evidence for
the density correlations discussed in the present paper. Our interpretation relies on the structural
property of the solutions to the Bogoliubov theory (\ref{bog}), but it proceeds by applying a
different physical reasoning which recognizes that the two detected particles are accompanied by
collision partners propagating in the opposite directions, and we hence observe part of a four-atom
state. This is an appealing picture, in particular because the prediction of the coincidence signal,
and in particular its width, relates to the transverse spreading of the pair wave functions of
oppositely propagating atoms after the bipartite collisions.

As we discussed in the text, when observed from only one direction, the reduced density matrix of the
expanding atomic cloud is similar to a thermal state. This density matrix is sufficient to predict the
outcome of any measurement on the observed part of the system, and it explains the experimental
findings as an analogue of the observed bunching of the photons from a thermal/chaotic light source.
The optical Hanbury Brown and Twiss experiment has a characteristic transverse spatial scale over
which the correlation falls to unity, related to the transverse momentum distribution of the photons
impinging on the detector, and in a similar manner we have a finite transverse coherence length in the
atomic scattering experiment. We discussed the isotropic s-wave scattering, with possible corrections
due to anisotropy of the colliding clouds and spatial phase matching. In addition, one may apply a
periodic background potential, which may alter the energy dependence on the momentum vector of moving
atoms, and hence modify the scattering profile \cite{nir,klausmod}, and with confinement to one
dimension, it may lead to highly selective population of specific momentum states with strong,
observable quantum correlations \cite{kmh,campbell}.

It is interesting to recall that a mixed quantum state, i.e., a density matrix, can always be formally
obtained as the reduced state of a larger quantum system which is in a pure state, and in particular
any thermal quantum state of a bosonic degree of freedom can be modelled by a pure squeezed state in a doubled tensor space.
This is known as the "thermofields" formulation \cite{thermofield}, and for
example the single mode thermal state (\ref{onemode}) can be obtained as the trace over one of the
modes of a non-degenerate two mode squeezed state, as obtained, e.g., from a non-degenerate optical
parametric oscillator (OPO),
\begin{equation} \label{twomode}
|\psi_{OPO}\rangle \propto \sum_n t^n |n,n\rangle.
\end{equation}
In our four-atom analysis, the apparently thermal state arriving at nearby detectors, is
precisely part of such a larger system. The advantage of this insight is that the spatial scale of the
extended state, in our case the probability distribution of the total momentum of scattered atoms,
directly yields the density correlations in the reduced density matrix.

Finally, if the collision occurs between two different bosonic
species, the same kind of correlations will occur for the density correlations of each species, but
not for the cross correlation, where the recoiling atoms are distinguishable, and hence do not
interfere. For collisions between bosons and fermions the situation is different. Electrons have been demonstrated to show anti-bunching related to their fermionic character
\cite{electrons}, and anti-bunching has also been demonstrated for neutral
fermionic atoms \cite{jeltes,bloch}. Collisions between a
Bose condensate and a degenerate Fermi gas, where all fermions initially occupy orthogonal states, but
where Pauli blocking forbids more than one atom ending up in the same final state should lead to observable anti-bunching effects in the scattered bosons.

\begin{acknowledgments}
The authors gratefully acknowledge discussions with Karen Kheruntsyan and financial support from the
European Union integrated project SCALA and the ESF cold atom network QUDEDIS. The Atom Optics group
of LCFIO is a member of the IFRAF institute, and is supported by the french ANR and by the Atom Chips
program of the European Union.
\end{acknowledgments}

\end{document}